\documentclass[
    reprint,
    amsmath,amssymb,
    aps,pre
]{revtex4-2}

\usepackage{graphicx}
\usepackage{bm}
\usepackage{amsfonts}
\usepackage{hyperref}
\usepackage{enumitem}
\usepackage{mathtools}

\begin{document}

\title{Collective action through adaptive awareness}

\author{Afolabi Ariwayo}
\author{Bryce Morsky}
\author{Malbor Asllani}
\email{malborasllani@gmail.com}

\affiliation{Department of Mathematics, Florida State University}

\date{\today}

\begin{abstract}
Collective actions emerge through the interplay between social influence and awareness. We introduce a nonlinear opinion-dynamics framework on networks in which social influence, shaped by individual interactions and community structure, promotes collective action, while awareness regulates the amount of reinforcement required for adoption. Using a degree-based mean-field reduction, we show that the competition between effective social influence and abandonment controls both the onset and persistence of collective action. Changes in awareness modify the nonlinear adoption mechanism itself, enabling populations to transition between highly responsive and weakly responsive collective states. This generates discontinuous transitions, bistability, and hysteresis, allowing collective action to persist even after the conditions that initially promoted it have weakened. We illustrate these effects through coupled disease-mitigation and resource-consumption dynamics, where external pressures act by reshaping awareness rather than social influence. More broadly, our results identify awareness as a fundamental link between environmental conditions, social interactions, and the emergence and persistence of collective action.
\end{abstract}

\maketitle

\emph{Introduction}---The emergence of collective behavior from interactions among individuals is a central problem in social science, statistical physics, and network science \cite{granovetter1983threshold,jackson2008social,lazer2009computational,castellano_statistical_2009,easley2010networks,newman_networks_2018}. Advances in computational social science and the increasing availability of large-scale behavioral data have enabled quantitative investigations of how opinions, information, and behaviors spread through networked populations \cite{centola_spread_2010,cinelli2021echo}. These approaches have been successfully applied to diverse phenomena, including cultural dissemination \cite{axelrod1997dissemination}, language evolution \cite{nowak2002computational}, collective human behavior \cite{barabasi2005origin}, and opinion formation \cite{bail2018exposure,tkavcik2016information}. A fundamental challenge across these systems is to understand how local interactions give rise to large-scale collective action.

Many models of social contagion assume that adoption results from pairwise influence between individuals. In practice, however, behavioral change adapts to external factors \cite{schultner2025feature}, often requiring reinforcement from multiple social contacts before adoption occurs \cite{centola2007complex}. Such reinforcement mechanisms introduce nonlinearities that can qualitatively alter collective dynamics, producing thresholds, multiple equilibria, abrupt transitions, and hysteresis \cite{liu_influence_1986,liu_dynamical_1987}. Understanding how social reinforcement shapes collective behavior is therefore essential for explaining the emergence and persistence of coordinated actions in human populations.

A key factor underlying reinforcement is awareness. Individuals differ in their responsiveness to social signals: some readily adopt behaviors after limited exposure, whereas others require repeated confirmation before acting. Awareness may itself evolve in response to external conditions, perceived risks, or environmental pressures, thereby changing the amount of reinforcement required for adoption. Despite its importance, awareness is typically incorporated indirectly through modifications of adoption, transmission, or recovery rates rather than through the reinforcement process itself.
This contrasts with related feedback mechanisms that instead rescale the input to an otherwise fixed-shape nonlinearity, whether through a linear trust weighting between social and environmental signals \cite{couthures2025bifurcation} or through gain modulation of a saturating response \cite{bizyaeva2022nonlinear}; here awareness instead reshapes the nonlinearity itself, fundamentally altering the adoption process rather than merely its drivers.

Awareness-driven behavioral responses play an important role in many real-world systems. During infectious disease outbreaks, for example, information exchange influences risk perception and the adoption of mitigation measures such as vaccination, mask usage, and social distancing \cite{eksin2017disease,paarporn2017networked}. As a consequence, epidemic spreading and behavioral adoption become coupled processes that influence one another through feedback \cite{10.1098/rspb.2003.2410,bauch2005imitation,bury2019charting,bauch2012evolutionary,morsky2023impact,morsky2025vaccination,eksin2019systematic,glaubitz2020oscillatory,saad2023dynamics}. Similar feedback mechanisms arise in socio-environmental systems, where collective actions respond to changing environmental conditions. These observations have motivated extensive research on coupled behavior--disease and awareness--epidemic dynamics \cite{wang2015coupled,granell2013dynamical,funk2010modelling,wu2012impact}.

Here we introduce a nonlinear opinion-dynamics framework on networks in which awareness directly regulates the level of social reinforcement required for adoption. Using a degree-based mean-field reduction \cite{pastor2001epidemic,barrat2008dynamical}, we show that collective action is governed by the competition between abandonment and an effective social-influence parameter that combines individual propensity for influence with amplification generated by community structure. This interaction gives rise to extinction, threshold, and bistable regimes, including discontinuous transitions and hysteresis. We further couple the opinion dynamics to epidemic and resource dynamics through awareness-mediated feedback, demonstrating how external pressures can reshape collective behavior by modifying awareness rather than social influence itself. More broadly, our results identify awareness as a fundamental link between environmental conditions, social interactions, and the emergence and persistence of collective action.


\emph{Awareness-regulated opinion dynamics}---We consider the spread of an opinion or collective action in a networked population. Individuals may adopt the action through social influence and later abandon it due to forgetting, fatigue, or changing perceptions, becoming susceptible to influence again \cite{pastor2001epidemic_2,smaldino2021coupled}. This repeated cycle of adoption and abandonment captures the dynamic nature of behavioral change in social systems.

Let $x_i(t)$ denote the fraction of individuals at node $i$ that have adopted the action, while $1-x_i(t)$ denotes the fraction that have not. A key ingredient of the model is awareness-mediated social reinforcement. In many social systems, a single interaction is insufficient to trigger adoption. Instead, individuals often require repeated confirmation from neighboring adopters before changing their behavior. We capture this through a nonlinear dependence of the
adoption rate on the fraction of adopters among an individual's neighboring contacts, $x_j^d$, summed over those neighbors $j$, where the exponent $d$ quantifies the amount of social reinforcement required for adoption. A related exponent appears in Ref.~\cite{chiba2024social}, where a power-law threshold distribution yields a factor $x_j^{\gamma}$ interpolating between simple ($\gamma=1$) and complex ($\gamma\neq1$) contagion. There $\gamma$ is fixed, whereas here $d$ is regulated by awareness through environmental feedback [see Eq.~\eqref{eq:d_feedback}], reshaping the bifurcation structure itself. For $d > 1$, adoption instead requires a threshold level of collective support before it
can take hold, giving rise to bistability between low- and high-adoption states. In this sense, $d$ captures the population's awareness or sensitivity to social influence, interpolating between overreactive and underreactive regimes.

Collective adoption is governed by two complementary factors. The first is the intrinsic influence strength $\beta$, which quantifies how effectively individuals persuade one another to adopt a given opinion or action. The second is the network structure, which determines how social interactions accumulate and reinforce adoption throughout the population. Consequently, collective behavior depends both on individual propensity for influence and on the opportunities for reinforcement provided by the social environment.

Assuming that influence propagates through a network with adjacency matrix $A$, the adoption dynamics at node $i$ are described by

\begin{equation}
\dot{x}_i
=
-\gamma x_i
+
\beta \left(1-x_i\right)
\sum_j A_{ij}x_j^{d},
\label{eq:orig_eq2}
\end{equation}
where $\gamma$ denotes the abandonment rate. The first term represents the loss of adopters, whereas the second term captures reinforcement-driven adoption from neighboring nodes. The adjacency matrix $A$ determines which social interactions are possible, while the nonlinear term $x_j^{d}$ accounts for the cumulative influence generated by multiple adopters.

To obtain analytical insight, we adopt a degree-based mean-field approximation developed in~\cite{qian2025explosive} and subsequently used in~\cite{qian2026mesoscopic}. Assuming a relatively narrow degree distribution and writing $x_i=\langle x\rangle+\delta x_i$ and $k_i=\langle k\rangle+\delta k_i$, where $k_i$ denotes the degree of node $i$, averaging Eq.~\eqref{eq:orig_eq2} over the network yields

\begin{equation}
\langle \dot{x}\rangle
=
-\gamma\langle x\rangle
+
\tilde{\beta}
\langle x\rangle^{d}
\left(1-\langle x\rangle\right),
\label{eq:MF}
\end{equation}
where $\tilde{\beta}=\beta\langle k\rangle$. Equation~\eqref{eq:MF} identifies $\tilde{\beta}$ as an effective social-influence parameter that combines individual propensity for influence $\beta$ with amplification generated by community structure $\langle k\rangle$. Thus, behavioral traits and social connectivity affect the emergence of collective action through the same control parameter. A detailed derivation is provided in the Appendix.

\begin{figure*}[t]
\centering
\includegraphics[width = \linewidth]{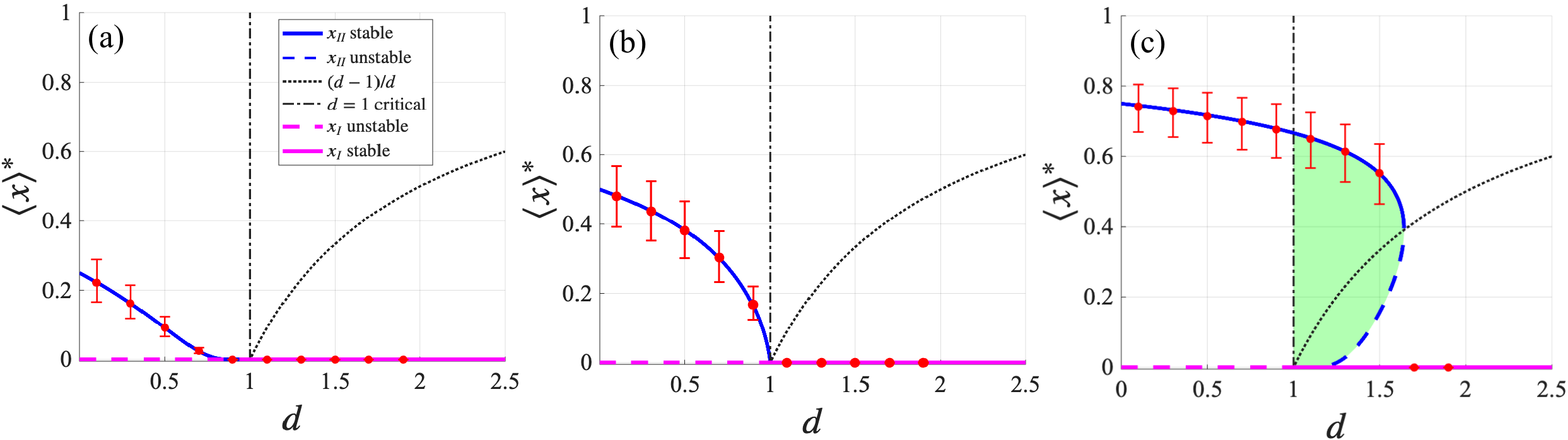}
\caption{Analytical bifurcation diagrams of the mean-field model for $(\gamma,\beta)=(30,0.25)$, $(30,0.74)$, and $(10,0.74)$ in panels (a)--(c), corresponding to the regimes $\gamma>\tilde{\beta}$, $\gamma=\tilde{\beta}$, and $\gamma<\tilde{\beta}$, respectively. These three cases represent extinction of collective action, the critical threshold separating adoption and abandonment, and a bistable regime characterized by hysteresis. In panel (c), the green shaded region indicates the basin of attraction of the nontrivial equilibrium, where the asymptotic outcome depends on both the initial condition and the direction of variation of the awareness parameter $d$. Solid and dashed curves denote stable and unstable equilibria, respectively. Red markers and error bars show the mean and range of equilibrium values obtained from simulations on an Erd\H{o}s--R\'enyi network with number of nodes $n=100$ and wiring probability $p=0.4$.}
\label{fig:combined_bifurcation}
\end{figure*}


\emph{Emergence and persistence of collective action}---Equation~\eqref{eq:MF} always admits the trivial equilibrium $\langle x\rangle^{*}=0$. For $0<d<1$, the linearization is singular at the origin and stability is assessed using Chetaev's theorem \cite{Khalil2002}. Consider the positive function
\(
V(\langle x\rangle)=\langle x\rangle^2/2,
\)
whose derivative along trajectories of Eq.~\eqref{eq:MF} is
\begin{equation}
\dot V(\langle x\rangle)
=
\langle x\rangle^2
\left[
-\gamma
+
\tilde{\beta}\langle x\rangle^{d-1}
(1-\langle x\rangle)
\right].
\end{equation}
Chetaev's theorem requires $\dot V>0$ in a punctured neighborhood of the origin. Defining $\varepsilon_c$ through
\begin{equation}
\frac{1-\varepsilon_c}{\varepsilon_c^{\,1-d}}
=
\frac{\gamma}{\tilde{\beta}},
\end{equation}
one has $\dot V>0$ for all $0<\varepsilon<\varepsilon_c$. Consequently, arbitrarily small levels of adoption are amplified, allowing collective action to emerge from weak social influence.

Besides the trivial equilibrium, nontrivial equilibria satisfy

\begin{equation}
\langle x\rangle^{*(d-1)}
\bigl(1-\langle x\rangle^{*}\bigr)
=
\frac{\gamma}{\tilde{\beta}}
\equiv c.
\label{eq:nonzero_eq}
\end{equation}
Defining
\begin{equation}
g(x)=x^{d-1}(1-x),
\end{equation}
the existence and stability of nontrivial equilibria are determined by the geometry of $g(x)$.
For $0<d<1$, the function $g(x)$ decreases monotonically from $+\infty$ to $0$ on $(0,1)$ (see Appendix). Consequently, Eq.~\eqref{eq:nonzero_eq} admits a unique nontrivial equilibrium for any $c>0$. Substituting Eq.~\eqref{eq:nonzero_eq} into the derivative of the vector field yields
\begin{equation}
f'(\langle x\rangle^{*})
=
\gamma
\left[
(d-1)
-
\frac{\langle x\rangle^{*}}
{1-\langle x\rangle^*}
\right].
\end{equation}
Therefore,
\begin{equation}
\langle x\rangle^*
>
\frac{d-1}{d}
\label{eq:d_cond}
\end{equation}
is the condition for stability of the nontrivial equilibrium. Hence, for $0<d<1$, the system possesses a unique stable adoption state while the trivial state is unstable.

At $d=1$, the stability of the trivial equilibrium is determined by
\begin{equation}
f'(0)=-\gamma+\tilde{\beta}.
\end{equation}
Thus, the origin is stable for $\gamma>\tilde{\beta}$ and unstable for $\gamma<\tilde{\beta}$. Evaluating the derivative at the nontrivial equilibrium yields
\begin{equation}
f'(\langle x\rangle^{*})
=
\gamma-\tilde{\beta},
\end{equation}
showing that the nontrivial equilibrium exists and is stable only when $\gamma<\tilde{\beta}$. Thus, $d=1$ marks the transition separating extinction from persistent collective action.

For $d>1$, evaluating the derivative of the vector field at the origin gives
\begin{equation}
f'(0)=-\gamma<0.
\end{equation}
Thus, the trivial equilibrium is stable for all $d>1$. The structure of the nontrivial equilibria is determined by the geometry of $g(x)$. For $d>1$, a straightforward calculation shows that the function $g(x)$ develops a maximum at
\begin{equation}
x_m=\frac{d-1}{d}.
\end{equation}
Consequently, Eq.~\eqref{eq:nonzero_eq} may admit zero, one, or two nontrivial solutions depending on the value of $c=\gamma/\tilde{\beta}$. The stability condition Eq.~\eqref{eq:d_cond} identifies the branch to the right of $x_m$ as stable and the branch to the left as unstable.
The critical ratio is inherited from the $d=1$ case, where the nontrivial equilibrium $\langle x\rangle^{*}=1-\gamma/\tilde{\beta}$ exists only for $\gamma<\tilde{\beta}$ and collides with the origin at $\gamma=\tilde{\beta}$. The resulting bifurcation structure therefore depends on the ratio $\gamma/\tilde{\beta}$. For $\gamma>\tilde{\beta}$, the nontrivial branch disappears at $d=1$ and the dynamics converge to the trivial state, corresponding to Fig.~\ref{fig:combined_bifurcation}(a). For $\gamma=\tilde{\beta}$, the nontrivial branch collides with the origin at $d=1$, producing the threshold configuration shown in Fig.~\ref{fig:combined_bifurcation}(b). Finally, for $\gamma<\tilde{\beta}$, a stable upper branch coexists with the stable trivial equilibrium and is separated from it by an unstable branch, generating the bistable and hysteretic regime shown in Fig.~\ref{fig:combined_bifurcation}(c). These three regimes determine whether collective action vanishes, emerges at threshold, or persists through bistability and hysteresis. Direct simulations on Erdős–Rényi networks confirm the predicted bifurcation structure; comparable agreement is found for scale-free and empirical network topologies (Supplemental Material).

\begin{figure*}[t]
\centering
\includegraphics[width = \linewidth]{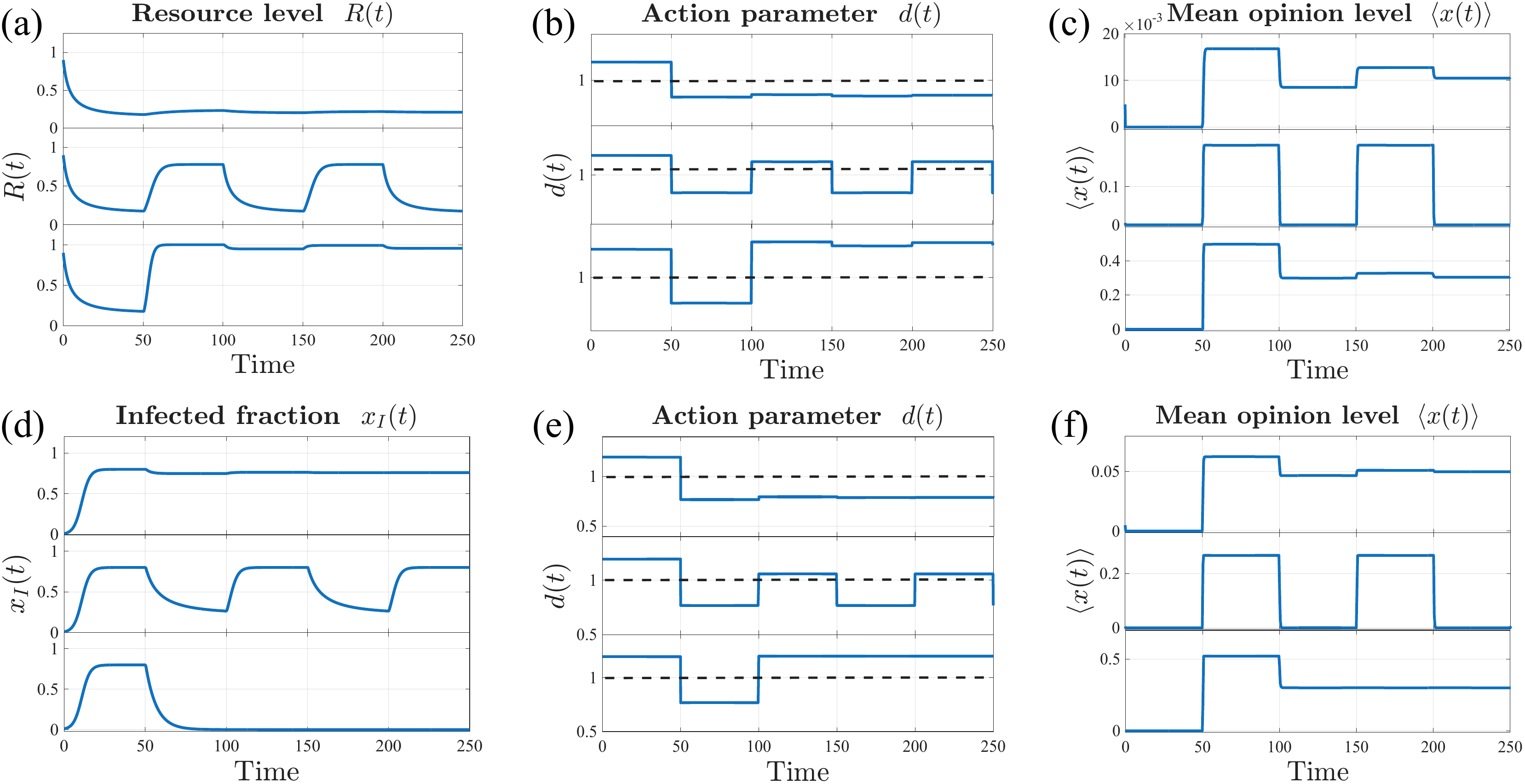}
\caption{
Panels (a)--(c) show the coupled opinion--resource system, whereas panels (d)--(f) show the coupled opinion--epidemic system. In each case, the left, middle, and right columns display the primary dynamical variable $R(t)$ or $x_I(t)$, the reinforcement parameter $d(t)$, and the mean opinion level $\langle x(t)\rangle$, respectively. From top to bottom, the rows correspond to $(\gamma,\beta)=(18,0.25)$, $(18,0.45)$, and $(10,0.45)$ for both the resource model, and the SIS model. When $\gamma>\tilde{\beta}$, forgetting dominates reinforcement and behavioral adoption remains weak, leading to resource depletion or persistent disease prevalence. Near $\gamma=\tilde{\beta}$, the competition between adoption and relaxation generates oscillatory dynamics. When $\gamma<\tilde{\beta}$, strong reinforcement sustains behavioral adoption, promoting resource recovery and suppressing disease transmission. Periodic updates of $d(t)$ from the resource level or disease prevalence produce the step-like transitions visible in the middle column. Parameters for the opinion--resource model are $\alpha_0=0.3$, $K=1$, $C_0=0.25$, $\eta_1=3$, $\eta_2=2$, $d_0=1.2$, and $\zeta=0.25$. Parameters for the opinion--epidemic model are $\beta_0=0.5$, $\gamma_0=0.1$, $\eta_1=2.5$, $\eta_2=1$, $d_0=1.2$, and $\zeta=0.45$. The resetting period is $T=50$ in both cases, and the network is that of Fig.~\ref{fig:combined_bifurcation}.}
\label{fig:coupled_dynamics_combined}
\end{figure*}


\emph{Coupled opinion--action dynamics}---To illustrate, in a minimal and intentionally schematic setting, how awareness-regulated opinion dynamics reshapes collective outcomes through feedback, we consider two representative examples: resource management and epidemic spreading. In both cases, the average opinion level $\langle x\rangle(t)$ evolves according to Eq.~\eqref{eq:MF}. The state of the system modifies the awareness level through periodic updates of the reinforcement exponent,

\begin{equation}
d=d_0\left[1-\zeta\, y(nT)\right],
\label{eq:d_feedback}
\end{equation}
where $y$ denotes the relevant system variable. Unlike conventional feedback mechanisms that modify dynamical rates, Eq.~\eqref{eq:d_feedback} acts directly on the awareness parameter $d$, changing the amount of social reinforcement required for adoption and thereby reshaping the underlying bifurcation structure. Through this mechanism, collective behavior modifies the underlying dynamics, while the resulting system state feeds back onto future adoption by regulating the responsiveness of individuals to social signals.

As a first example, we consider a renewable resource governed by

\begin{equation}
\dot{R}
=
\alpha_RR\left(1-\frac{R}{K}\right)
-
C_RR,
\label{eq:logistic_harvesting}
\end{equation}
where $R(t)$ is the resource level and $K$ is the carrying capacity. Behavioral adoption modifies harvesting and regeneration according to

\begin{equation}
C_R=C_0(1-\eta_1\langle x\rangle),
\qquad
\alpha_R=\alpha_0(1+\eta_2\langle x\rangle).
\end{equation}

Here, $C_0$ and $\alpha_0$ denote the baseline harvesting and growth rates. Consequently, increasing adoption reduces exploitation while enhancing resource recovery. In Eq.~\eqref{eq:d_feedback}, the feedback variable is identified with the resource level, $y\equiv R$.

Second, we consider the SIS epidemic model,

\begin{equation}
\dot{x}_I
=
-\gamma_Ix_I
+
\beta_I(1-x_I)x_I,
\label{eq:sis}
\end{equation}
where $x_I(t)$ denotes the infected fraction. Behavioral adoption modifies transmission and recovery as

\begin{equation}
\beta_I=\beta_0(1-\eta_1\langle x\rangle),
\qquad
\gamma_I=\gamma_0(1+\eta_2\langle x\rangle).
\end{equation}
Here, $\beta_0$ and $\gamma_0$ are the transmission and recovery rates. Thus, increasing adoption suppresses transmission while enhancing recovery. In this case, the feedback variable entering Eq.~\eqref{eq:d_feedback} is the disease prevalence, $y\equiv x_I$.

Figure~\ref{fig:coupled_dynamics_combined} illustrates the resulting dynamics. Although resource management and epidemic mitigation represent distinct processes, both are governed by the same feedback mechanism: collective behavior modifies the underlying dynamics, while the resulting system state regulates awareness through Eq.~\eqref{eq:d_feedback}. Depending on the balance between social influence, awareness, and abandonment, this interplay generates stable equilibria, oscillatory dynamics, bistability, and hysteresis.


\emph{Discussion}---We introduced a nonlinear opinion-dynamics framework in which collective action emerges through the interplay between social influence and awareness. A degree-based mean-field reduction reveals that the dynamics are governed by the competition between abandonment, represented by $\gamma$, and the effective social influence $\tilde{\beta}=\beta\langle k\rangle$, which combines individual propensity for influence with amplification generated by community structure. This competition determines whether collective action vanishes, emerges continuously, or persists through bistability and hysteresis.
The central ingredient of the framework is the awareness parameter $d$, which regulates the amount of social reinforcement required for adoption. Unlike conventional feedback mechanisms that modify adoption or transmission rates, awareness directly alters the nonlinear adoption process itself, thereby changing the underlying bifurcation structure. As a result, external conditions influence collective behavior not only by shifting equilibrium levels but also by reshaping the phase portrait governing the emergence and persistence of adoption.
The coupled resource and epidemic examples illustrate this mechanism in two distinct settings. In both cases, the system state feeds back onto awareness, which in turn regulates the responsiveness of individuals to social influence. This feedback can sustain collective action long after the conditions that initially promoted it have weakened, generating memory effects, bistability, and hysteresis. More broadly, our results identify awareness as a fundamental link between environmental conditions, social interactions, and collective behavior, providing a framework for understanding how external pressures reshape the emergence and persistence of coordinated action in networked populations.

\bibliographystyle{apsrev4-2}
\bibliography{citations}

\appendix
\begin{appendix}

\section{Degree-Based Mean-Field Approximation}
\label{sec:DBMF}

Following the Degree-Based Mean-Field (DBMF) framework \cite{pastor2015epidemic}, we assume that the degree fluctuations are small compared to the mean degree,
\begin{equation*}
\frac{\delta k_i}{\langle k\rangle}\ll 1,
\qquad
\delta k_i=k_i-\langle k\rangle .
\end{equation*}
Equivalently, the degree distribution is assumed to be sufficiently concentrated around its mean in a relative sense. Under this assumption, we further assume that the local adoption level is correlated with the node degree,
\begin{equation*}
x_i(t)\sim k_i,
\end{equation*}
where $k_i$ denotes the degree of node $i$.

Let
\begin{equation*}
\langle x\rangle=\frac{1}{n}\sum_i x_i,
\qquad
\langle k\rangle=\frac{1}{n}\sum_i k_i .
\end{equation*}
Averaging Eq.~\eqref{eq:orig_eq2} over all nodes yields
\begin{align}
\langle\dot{x}\rangle
&=
-\gamma\langle x\rangle
+
\frac{\beta}{n}
\sum_{i,j}
A_{ij}x_j^{d}
-
\frac{\beta}{n}
\sum_{i,j}
A_{ij}x_i x_j^{d},
\nonumber\\
&=
-\gamma\langle x\rangle
+
\frac{\beta}{n}
\sum_j k_j x_j^{d}
-
\frac{\beta}{n}
\sum_{i,j}
A_{ij}x_i x_j^{d}.
\label{eq:dbmf1}
\end{align}
Writing
\begin{equation*}
x_i=\langle x\rangle+\delta x_i,
\qquad
k_i=\langle k\rangle+\delta k_i ,
\end{equation*}
and substituting into Eq.~\eqref{eq:dbmf1} gives
\begin{align*}
\langle\dot{x}\rangle
&=
-\gamma\langle x\rangle
+
\frac{\beta}{n}
\sum_j
(\langle k\rangle+\delta k_j)
(\langle x\rangle+\delta x_j)^d
\nonumber\\
&\quad
-
\frac{\beta}{n}
\sum_{i,j}
A_{ij}
\left(\langle x\rangle+\delta x_i\right)
\left(\langle x\rangle+\delta x_j\right)^d .
\end{align*}
To first order in the fluctuations,
\begin{equation*}
(\langle x\rangle+\delta x_j)^d
\approx
\langle x\rangle^d
+
d\langle x\rangle^{d-1}\delta x_j .
\end{equation*}
Substituting this approximation into Eq.~\eqref{eq:dbmf1} and expanding yields
\begin{align}
\langle\dot{x}\rangle
&=
-\gamma\langle x\rangle
+
\beta\frac{1}{n}
\sum_j
\langle k\rangle\langle x\rangle^d
+
\beta\frac{1}{n}
\sum_j
\delta k_j\langle x\rangle^d
\nonumber\\
&\quad
+
\beta\frac{d}{n}
\sum_j
\langle k\rangle
\langle x\rangle^{d-1}\delta x_j
+
\beta\frac{d}{n}
\sum_j
\delta k_j
\langle x\rangle^{d-1}\delta x_j
\nonumber\\
&\quad
-
\frac{\beta}{n}
\sum_{i,j}
A_{ij}
\langle x\rangle^{d+1}
-
\frac{\beta d}{n}
\sum_{i,j}
A_{ij}
\langle x\rangle^{d}\delta x_j
\nonumber\\
&\quad
-
\frac{\beta}{n}
\sum_{i,j}
A_{ij}
\langle x\rangle^{d}\delta x_i
-
\frac{\beta d}{n}
\sum_{i,j}
A_{ij}
\langle x\rangle^{d-1}
\delta x_i\delta x_j .
\label{eq:dbmf_expand}
\end{align}
Using
\begin{equation*}
\sum_i \delta x_i=0,
\qquad
\sum_i \delta k_i=0,
\end{equation*}
and neglecting second-order fluctuation terms, namely terms involving $\delta k_i\delta x_j$ and $\delta x_i\delta x_j$, we obtain
\begin{equation*}
\langle\dot{x}\rangle
=
-\gamma\langle x\rangle
+
\beta\langle k\rangle\langle x\rangle^{d}
-
\beta\langle k\rangle\langle x\rangle^{d+1}.
\end{equation*}
Defining $\tilde{\beta}=\beta\langle k\rangle$, the dynamics reduce to
\begin{equation}
\langle\dot{x}\rangle
=
-\gamma\langle x\rangle
+
\tilde{\beta}
\langle x\rangle^{d}
\bigl(1-\langle x\rangle\bigr),
\label{eq:avg5}
\end{equation}
which corresponds to Eq.~\eqref{eq:MF} in the main text.


\section{Chetaev Instability Criterion}
\label{sec:chetaev}

For completeness, we state the version of Chetaev's theorem used in the stability analysis of the trivial equilibrium. Consider

\begin{equation}
\dot{x}=f(x),
\end{equation}
with an equilibrium at the origin. If there exists a continuously differentiable function $V(x)$ such that

\begin{enumerate}
\item $V(x)>0$ in a domain whose boundary contains the origin,
\item $\dot V(x)>0$ throughout that domain,
\end{enumerate}
then the origin is an unstable equilibrium \cite{Khalil2002}.

\section{Analysis of the Nontrivial Equilibrium}
\label{sec:convex}

The nontrivial equilibria of Eq.~\eqref{eq:MF} satisfy

\begin{equation}
x^{d-1}(1-x)=c,
\qquad
c=\frac{\gamma}{\tilde{\beta}},
\label{eq:C_nontrivial}
\end{equation}
with $x\in(0,1)$.
Defining

\begin{equation*}
F(d,x)=x^{d-1}(1-x)-c,
\end{equation*}
the nontrivial equilibrium branch is determined by

\begin{equation*}
F(d,x)=0.
\end{equation*}

\subsection{Existence and uniqueness for $d<1$}

Differentiating with respect to $x$ gives

\begin{equation}
F_x(d,x)
=
x^{d-2}
\big[(d-1)-dx\big].
\end{equation}
For $d<1$ and $x\in(0,1)$,

\begin{equation*}
(d-1)-dx<0,
\end{equation*}
and therefore

\begin{equation*}
F_x(d,x)<0.
\end{equation*}
Consequently, the implicit function theorem guarantees the existence of a unique smooth branch $x(d)$ satisfying Eq.~\eqref{eq:C_nontrivial}.

\subsection{Monotonicity}

Differentiating $F(d,x(d))=0$ yields

\begin{equation}
x'(d)
=
-\frac{F_d}{F_x}.
\end{equation}
Using

\begin{equation*}
F_d
=
x^{d-1}(1-x)\ln x,
\end{equation*}
and Eq.~\eqref{eq:C_nontrivial}, we obtain

\begin{equation}
x'(d)
=
-\frac{x(1-x)\ln x}
{(d-1)-dx}.
\label{eq:C_first}
\end{equation}
Since $\ln x<0$ and $(d-1)-dx<0$ for $d<1$, it follows

\begin{equation*}
x'(d)<0.
\end{equation*}
Hence the nontrivial equilibrium decreases monotonically as the reinforcement exponent increases.

\subsection{Convexity}
Differentiating Eq.~\eqref{eq:C_first} once more gives

\begin{equation}
x''(d)
=
-\frac{
F_{dd}
+
2F_{dx}x'
+
F_{xx}(x')^2
}
{F_x},
\end{equation}
where

\begin{align*}
F_{dd}
&=
x^{d-1}(1-x)(\ln x)^2,
\\
F_{dx}
&=
x^{d-2}
\Big(
[(d-1)(1-x)-x]\ln x
+
(1-x)
\Big),
\\
F_{xx}
&=
x^{d-3}
\Big(
(d-2)\big[(d-1)-dx\big]-d
\Big).
\end{align*}

\subsection{Behavior near $d=1$}

The qualitative behavior of the nontrivial equilibrium depends on the value of
\(
c=\gamma/\tilde{\beta}.
\)

\paragraph{Case $0<c<1$.}

At $d=1$, Eq.~\eqref{eq:C_nontrivial} admits the finite solution

\begin{equation*}
x(1)=1-c.
\end{equation*}
Since $0<c<1$, the equilibrium remains strictly positive at
$d=1$. Evaluating Eq.~\eqref{eq:C_first} at $d=1$ gives

\begin{equation*}
x'(1)
=
\frac{c(1-c)\ln(1-c)}
{1-c}
=
c\ln(1-c).
\end{equation*}
Since $\ln(1-c)<0$, it follows that

\begin{equation*}
x'(1)<0.
\end{equation*}
Thus, the nontrivial equilibrium decreases smoothly as $d$
increases through $d=1$. Expanding around $d=1$ yields

\begin{equation*}
x(d)
=
1-c
+
c\ln(1-c)(d-1)
+
\mathcal O\!\left((d-1)^2\right).
\end{equation*}

\paragraph{Case $c=1$.}

For $c=1$, Eq.~\eqref{eq:C_nontrivial} becomes

\begin{equation*}
x^{d-1}(1-x)=1.
\end{equation*}
The nontrivial branch collides with the origin at $d=1$,
since

\begin{equation*}
x(1)=0.
\end{equation*}
To determine the local scaling, we take logarithms,

\begin{equation*}
(d-1)\ln x+\ln(1-x)=0.
\end{equation*}
For $x\ll1$,

\begin{equation*}
\ln(1-x)\approx -x,
\end{equation*}
and therefore

\begin{equation*}
(d-1)\ln x \approx x.
\end{equation*}
Hence the nontrivial equilibrium approaches the origin
continuously as $d\to1^{-}$.

\paragraph{Case $c>1$.}

For $c>1$, no finite positive solution exists at $d=1$.
As $d\to1^{-}$, the nontrivial equilibrium approaches the
origin. Taking logarithms of Eq.~\eqref{eq:C_nontrivial} 

\begin{equation*}
(d-1)\ln x+\ln(1-x)=\ln c.
\end{equation*}
Since $x\to0$, we use

\begin{equation*}
\ln(1-x)=\mathcal O(x),
\end{equation*}
which yields the leading-order balance

\begin{equation*}
\ln x
\sim
\frac{\ln c}{d-1},
\end{equation*}
or equivalently,

\begin{equation*}
x(d)
\sim
\exp\!\left(
-\frac{\ln c}{1-d}
\right).
\label{eq:C_asymptotic}
\end{equation*}

Thus, the nontrivial equilibrium vanishes exponentially fast as
$d\to1^{-}$.

\end{appendix}

\end{document}